\documentclass[%
 reprint,
superscriptaddress,
 amsmath,amssymb,
 aps,
pra,
floatfix,
]{revtex4-2}

\usepackage{caption}
\usepackage{subcaption}
\usepackage[version=4]{mhchem}  
\usepackage{graphicx}
\usepackage{xcolor}
\usepackage{graphicx}
\usepackage{dcolumn}
\usepackage{bm}
\usepackage{ulem}

\DeclareRobustCommand{\erase}{\bgroup\markoverwith{\textcolor{red}{\rule[.5ex]{2pt}{0.4pt}}}\ULon}

\begin{document}

\preprint{APS/123-QED}

\title{Designing rotational motion of charges on plasmonic nanostructures\\ excited by circularly polarized light}

\author{Naoki Ichiji*}
\affiliation{Institute of Industrial Science, The University of Tokyo, 4-6-1 Komaba, Meguro-Ku, Tokyo 153-8505, Japan}
\email{ichiji@iis.u-tokyo.ac.jp}
\author{Takuya Ishida}
\affiliation{Institute of Industrial Science, The University of Tokyo, 4-6-1 Komaba, Meguro-Ku, Tokyo 153-8505, Japan}
\author{Ikki Morichika}
\affiliation{Institute of Industrial Science, The University of Tokyo, 4-6-1 Komaba, Meguro-Ku, Tokyo 153-8505, Japan}
\author{Daigo Oue}
\affiliation{Instituto de Telecomunica\c{c}\~{o}es, Instituto Superior T\'{e}cnico, University of Lisbon, 1049-001 Lisbon, Portugal}
\affiliation{The Blackett Laboratory, Imperial College London, London SW7 2AZ, United Kingdom}
\author{Tetsu Tatsuma}
\affiliation{Institute of Industrial Science, The University of Tokyo, 4-6-1 Komaba, Meguro-Ku, Tokyo 153-8505, Japan}
\author{Satoshi Ashihara}
\affiliation{Institute of Industrial Science, The University of Tokyo, 4-6-1 Komaba, Meguro-Ku, Tokyo 153-8505, Japan}


\begin{abstract}
Rotational motion of charges in plasmonic nanostructures plays an important role in transferring angular momentum between light and matter on the nanometer scale. Although sophisticated control of rotational charge motion has been achieved using spatially structured light, its extension to simultaneous excitation of the same charge motion in multiple nanostructures is not straightforward.
In this study, we perform model calculations to show that spatially homogeneous circularly polarized (CP) light can excite rotational charge motions with a high degrees of freedom by exploiting the rotational symmetry of the plasmonic structure and that of the plasmon mode. Finite-difference time-domain simulations demonstrate selective excitation of rotational charge motion for both isolated nanoplates and periodic array structures, showing that complex charge rotations can be manipulated by plane CP waves in a wide range of plasmonic structures.
\end{abstract}

\maketitle

\section{introduction}
The angular momentum transfer between light and matter has been extensively studied for more than a century. Various attempts have been made to transfer the angular momentum of light, including the spin angular momentum (SAM) or orbital angular momentum (OAM), to materials~\cite{Friese98Nature,Stilgoe22NatP}.
These attempts include the induction of spinning and orbital motions of nanoparticles~\cite{Zhao09OE,Monteiro18PRA,Fujiwara21NL,Karpinski22AOM,Wang11NC}, fabrication of chiral nanostructures~\cite{Saito18NL,Toyoda13PRL,Shimomura20APL,Ishida23APL}, and mass transfer in polymers~\cite{Ambrosio12NatC,Tomita23OE}.
Angular momentum transfer also causes various interesting phenomena related to the angular momentum of light, such as circularly polarized light scattering~\cite{Negoro23NL}, higher harmonic generation with handedness inversion~\cite{Konishi20Optica}, and enhancement of inverse Faraday effect~\cite{Alma24NanoP,Parchenko24APR, Cheng20NatP}.

\begin{figure}[b!]
  \begin{center}
  \includegraphics[width=8.6cm]{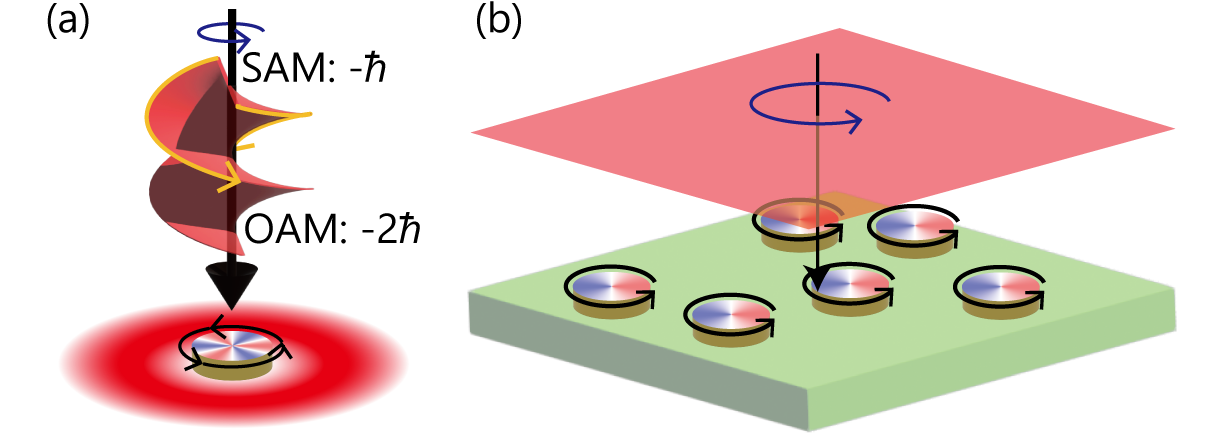}
  \end{center}
  \caption{Schematics of the rotational motion of the charge density on the plasmonic structures excited by (a) vortex beam and (b) spatially homogeneous circular polarized light. Black arrows indicate the charge rotation per oscillation period.}
  \label{Fig:image}
\end{figure}

Recently, several studies demonstrated that structured light with a designed spatial phase distribution can directly transfer their angular momentum to the rotational charge motion in plasmonic structures~\cite{Sakai16SciRep, Arikawa20SciAd, YangNanoscale23, Tanaka23LPR}.
For example, a vortex beam with $-2\hbar$ OAM and $-\hbar$ SAM can transfer a total of $-3\hbar$ angular momentum to plasmonic structure. This results in their charges rotating by $2\pi/3$ in one oscillation period (Figure~\ref{Fig:image} (a)).

Although direct transfer of angular momentum by structured light enables the control of charge motion with a high degree of freedom, this transfer typically occurs around the phase singularity—where the intensity is minimal.
This limitation restricts both the excitation efficiency of the induced charge motion and the number of objects that can be excited.
To simultaneously and uniformly excite the same rotational motion in numerous structures—ideal for observing and manipulating physical phenomena such as induced magnetic fields~\cite{Cheng22NL,Cheng20NatP}, the rotational Doppler effect~\cite{Pan19PRL}, and evanescent fields with SAM~\cite{Bliokh14NC}—excitation with spatially homogeneous light is desirable.

Spatially homogeneous, circularly polarized (CP) light may have great potential for simultaneously exciting rotational motion in numerous structures~\cite{Alma24NanoP}.
An inherent limitation of homogeneous CP light in controlling rotational charge motion is that the angular momentum of the CP light can attain one of only two states: right or left-handed.
Usually, the rotational charge motion induced by the CP light intuitively follows the rotation of the electric field, either right- or left-handed, which renders the altering of the direction or speed of the rotational motion challenging (Figure~\ref{Fig:image} (b)).
However, a numerical calculation on the torque acting on plasmonic structures induced by CP light has showed an intriguing phenomenon: charges on a plasmonic nanoplate excited by plane CP light rotate in different directions, depending on the rotational symmetry of the plate and the order of plasmonic resonance~\cite{Lee14NanoP}.
The fact that the motion of charges excited by CP waves deviates from the rotation of the incident electric field suggests that the rotational charge motion can be manipulated with a high degree of freedom by a plane CP wave through the geometric design of the plasmonic structure.

For the advanced control of rotational charge motion using CP light, it is essential to accurately grasp how the geometry of the plasmonic structure affects charge motions.
In this study, we performed model calculations to show how the rotational symmetry of the plasmonic structure and that of the plasmon mode affect the rotational charge motion.
We found that the proper selection of these two rotational symmetries enables the selective excitation of complex rotational charge motions with homogeneous CP light.
The analytically obtained insight in this study can be generally applicable to versatile systems that have rotational symmetry, including regular nanoplates.
Finite-difference time-domain (FDTD) simulations validated the charge behavior predicted by the analytical model for both single nanoplates and periodically arranged hole structures. These results suggest that the rotational charge motion could be manipulated by plane CP light in a wide range of plasmonic structures.

\section{Model Calculations on Azimuthally Periodic Systems}
\subsection{$C_{4}$ symmetry}
Model calculations were performed to predict the rotational motion of charges in plasmonic nanostructures with different rotational symmetries. First, we consider a plasmonic system with $C_{4}$ symmetry and its azimuthal periodic plasmon modes.
The charge density for the azimuthal periodic plasmon mode excited by $x$-polarized linearly polarized (LP) light, $\rho^{x\mathrm{pol}}_{m}(r,\theta)$, is expressed in polar coordinates using radius $r$ and angle $\theta$ as follows.
\begin{align}
   \rho^{x\mathrm{pol}}_{m}(r,\theta)\!&=\!A(r)\cos[m\theta],\label{Eq:xpol}
\end{align}
where $A(r)$ denotes the radial distribution.
The variable $m$ is a positive integer that represents the order of the rotational symmetry of the charge density, where $m\!=\!1, 2, $ and 3 correspond to the dipole, quadrupole, and hexapole modes, respectively~\cite{Arikawa20SciAd}.

\begin{figure}[t]
  \begin{center}
  \includegraphics[width=8.6cm]{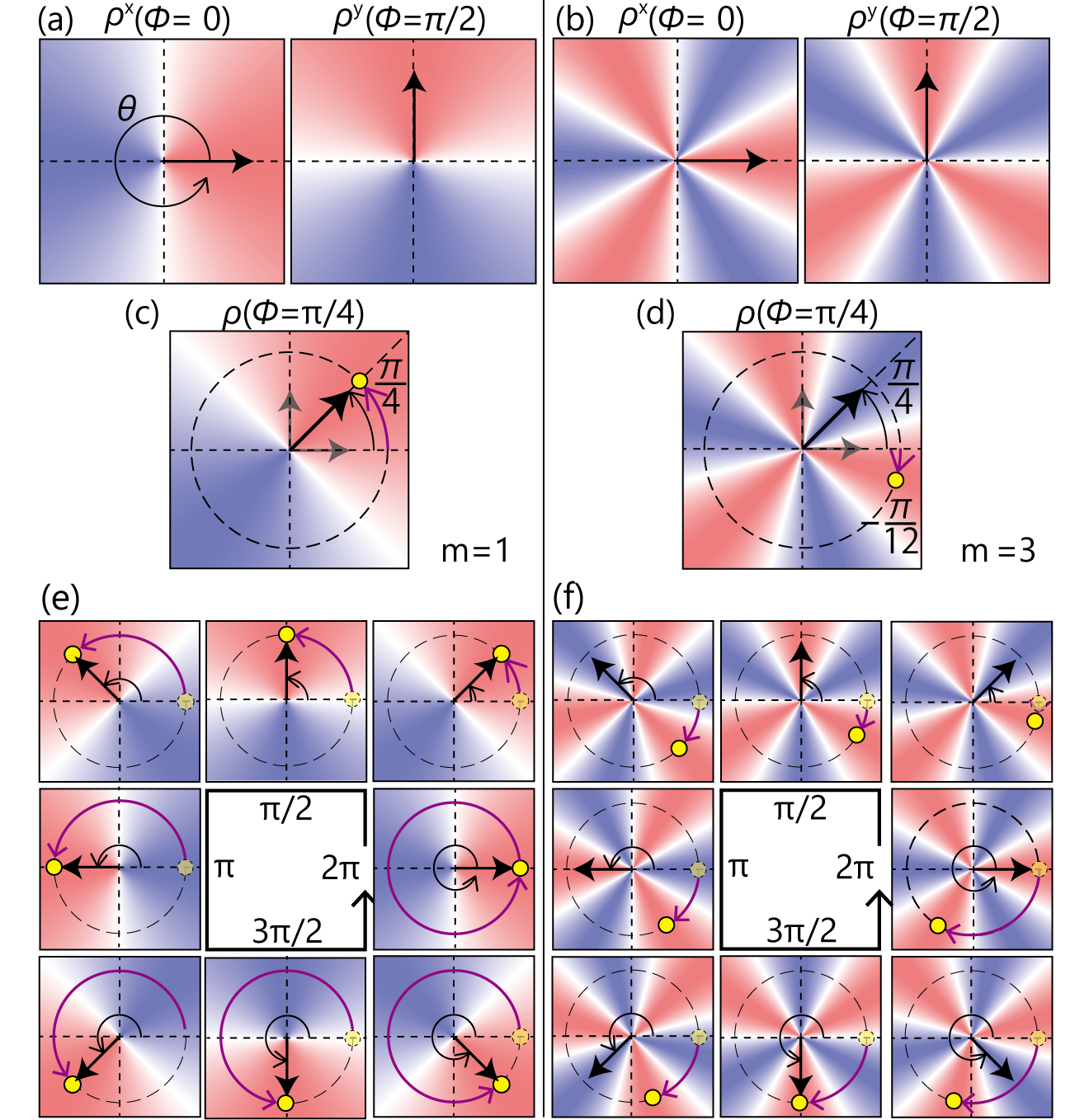}
  \end{center}
  \caption{(a,b) Simplified charge density distribution of the (a) dipole and (b) hexapole mode excited by $x$-and $y$-polarized linearly polarized (LP) light, respectively. The black arrows indicate the direction of the electric field of excitation light. (c and d) Charge distributions with $\phi\!=\!\pi/4$, calculated from the charge density of each of the $x$- and $y$-polarized excitation. The yellow circular markers and purple arrows indicate the angular shift where the charge density is maximum. (e and f) Angular distribution of charge density under circularly polarized excitation for one oscillation period.
}
  \label{Fig:model}
\end{figure}

In a system with $C_{4}$ symmetry, the charge density excited by $y$-polarized LP light, $\rho^{y\mathrm{pol}}_{m}(r,\theta)$, is required to be exactly equal to $\rho_{m}^{x\mathrm{pol}}$ rotated by $\pi/2$. This can be expressed as
\begin{align}
   \rho^{y\mathrm{pol}}_{m}(r,\theta)\!&=\rho^{x\mathrm{pol}}_{m}(r,\theta-\pi/2)\nonumber\\
   &=\!A(r)\cos[m\left(\theta-\pi/2\right)].\label{Eq:ypol}
\end{align}
By classifying the odd-order resonance modes into $m\!=\!(4j\pm 1)$ using integer $j$, we can divide Eq.~\ref{Eq:ypol} as follows:
\begin{equation}
\rho^{y\mathrm{pol}}_{m\!=4j\pm1}(r,\theta)\!=\!\pm A(r) \sin[m\theta]\label{Eq:case}.
\end{equation}
Note that even-order modes are forbidden in $C_{4}$ symmetric systems.
The modeled $\rho_{m}^{x\mathrm{pol}}$ and $\rho_{m}^{y\mathrm{pol}}$ for the dipole mode, $m\!=\!1\!=\!4\cdot0+1$ ($j\!=\!0$), and those of the hexapole mode, $m\!=\!3\!=\!4\cdot 1-1$ ($j\!=\!1$), are plotted in Figures~\ref{Fig:model} (a) and (b), respectively.
$A(r)$ is set to 1 for simplicity.

The charge density distribution excited by the LP light with a polarization angle $\phi$, $\rho_{m}(r,\theta;\phi)$, is expressed as a superposition of the charge distribution excited by $x$-polarized light and that excited by $y$-polarized light,
\begin{align}
\rho_{m}(r,\theta;\phi)\!&=\!\rho^{x\mathrm{pol}}_{m}(r,\theta)\cos[\phi]+\rho^{y\mathrm{pol}}_{m}(r,\theta)\sin[\phi]\label{Eq:phi}.
\end{align}
From Eqs.~\ref{Eq:xpol}-\ref{Eq:phi}, the charge density distribution of order $m\!=\!4j\pm1$ can be transformed as follows.
\begin{align}
&\rho_{m\!=\!4j\pm1}(r,\theta;\phi) \notag\\
&=\!A(r)\left(\cos[m\theta]\cos[\phi]\pm\sin[m\theta]\sin[\phi]\right) \notag\\
&=\!A(r)\cos\left[m\left(\theta\mp\frac{\phi}{m}\right)\right]\!=\!\rho^{x\mathrm{pol}}_{m}\left(r, \theta\mp\frac{\phi}{m}\right).
\label{Eq:add}
\end{align}
Equation~\ref{Eq:add} indicates that the charge distribution $\rho_{m}(r,\theta;\phi)$ is equal to $\rho^{x\mathrm{pol}}_{m}(r,\theta)$ rotated by $\phi/m$ counterclockwise (clockwise) for $m\!=\!4j+1~(m\!=\!4j-1)$.

In Figures~\ref{Fig:model} (c) and (d), we plot the calculated charge density distribution excited by LP light with polarization angle $\phi\!=\!\pi/4$, $\rho_{1}(\theta;\phi\!=\!\pi/4)$ and $\rho_{3}(\theta;\phi\!=\!\pi/4)$.
In the dipole mode plotted in Figure~\ref{Fig:model} (c), the angle at which the charge density exhibits its maximum value is oriented $\pi/4$, aligning with the direction of the electric field.
Conversely, in the hexapole mode, the maximum charge density shifts in the opposite direction to the shift of the electric field by $-\pi/12$.

It is noteworthy that the the charge density beneath the black arrow (aligned with the electric field vector) is positive for both horizontal and vertical polarizations (Figure~\ref{Fig:model} (b)), but turns negative for diagonal polarization (Figure~\ref{Fig:model} (d)).
This non-intuitive charge distribution cannot occur in systems with $C_{8}$ symmetry, such as regular octagons and isotropic disks, where $\rho(\theta-\pi/4)$ must be $\rho(\theta)$ rotated closkwise by $\pi/4$.
$C_4$-symmetric systems lack $C_8$ symmetry, resulting in horizontal and diagonal excitations producing fundamentally distinct outcomes (i.e., $\rho(\theta-\pi/4)$ differs from $\rho(\theta)$ rotated clockwise by $\pi/4$).

Under CP light irradiation, $\phi$ continuously varies from 0 to $\pm2\pi$ within an oscillation period.
Figures~\ref{Fig:model} (e) and (f) show the time evolution of the angular charge density distribution for modes $m\!=\!1$ and 3, over one oscillation period ($\phi$ varies continuously from 0 to $2\pi$), corresponding to the incidence of left-handed CP light.
For the dipole mode, the angle at which the charge density attains its maximum value, indicated by the yellow marker and purple arrow, follows the rotation of the electric field and rotates 2$\pi$ per oscillation period.
In contrast, the angle of maximum charge density in the hexapole mode rotates in the direction opposite to the rotation of the electric field, with a rotation angle of $2\pi/3$ per oscillation period.

\subsection{$C_{N}$ symmetric system}
\begin{figure}[t]
  \begin{center}
  \includegraphics[width=8.6cm]{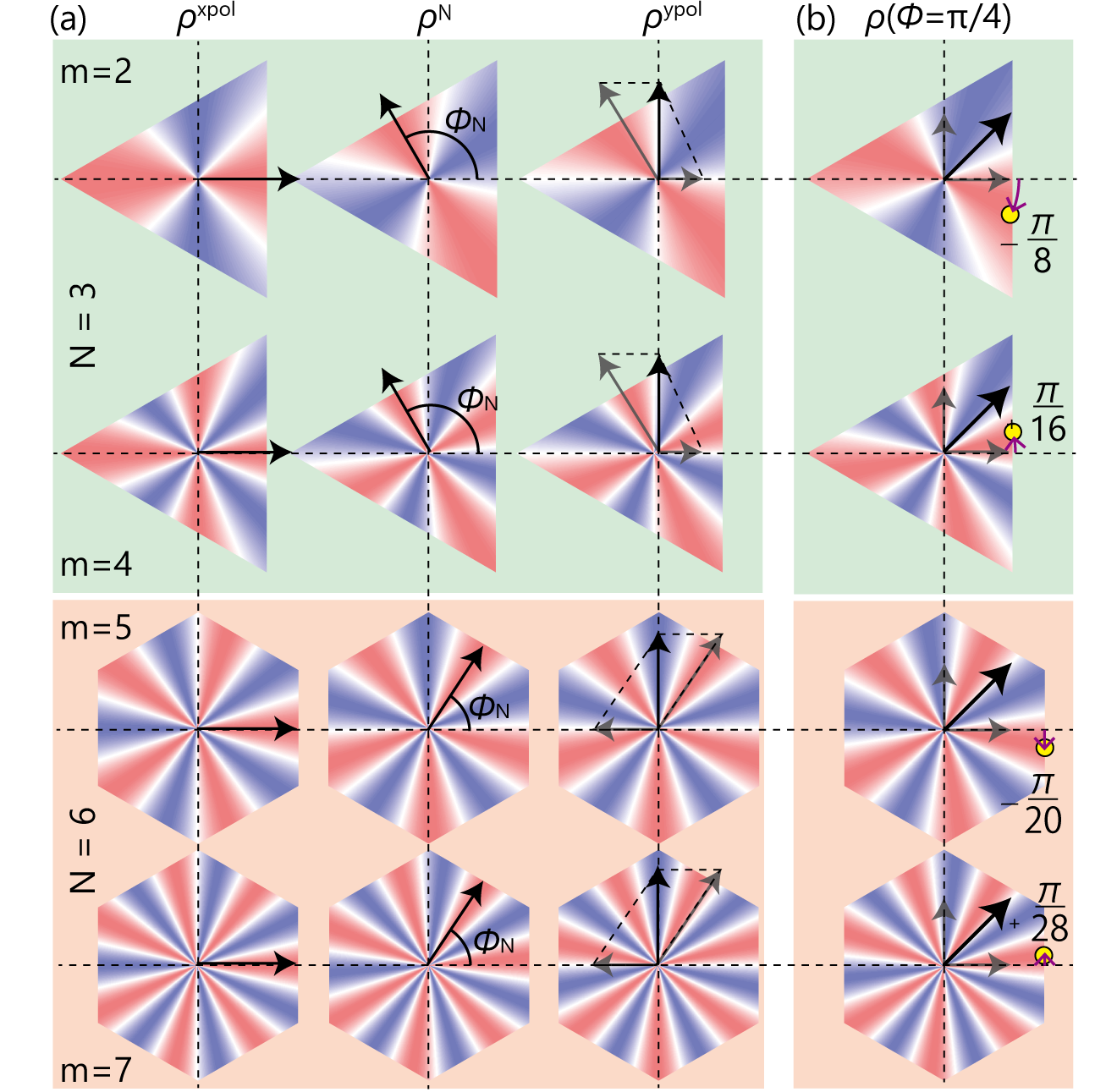}
  \end{center}
  \caption{(a) Calculated charge density distributions in the $C_{3}$ (top panel) and $C_{6}$ (bottom panel) symmetric system. From the left to right column, the polarization of the excitation lights is $\phi\!=\!0$, $\phi\!=\!\phi_{N}\!=\!2\pi/N$, and $\phi\!=\!\pi/2$.
  (b) Calculated charge density distributions for $\phi\!=\!\pi/4$. The yellow markers and purple arrows indicate the angular shift where the charge density is maximum.}
  \label{Fig:general}
\end{figure}
We now extend our discussion to general rotationally symmetric systems.
The $C_{N}$ rotationally symmetric systems require the following symmetry, using $\phi_{N}\!=\!2\pi/N$,
\begin{align}
   \rho^{N}_{m}(r,\theta; \phi_{N})\!&=\rho^{x\mathrm{pol}}_{m}\left(r,\theta-\phi_{N}\right)\!=\!A(r)\cos\left[m\left(\theta-\phi_{N}\right)\right],\label{Eq:ppol}
\end{align}
where, $\rho^{N}_{m}(r,\theta;\phi_{N})$ represents the charge density excited by the LP light with polarization angle $\phi_{N}$.
From Eq.~\ref{Eq:phi}, $\rho^{y\mathrm{pol}}_{m}(r,\theta)$ can be expressed as
\begin{align}
\rho_{m}^{y\mathrm{pol}}(r,\theta) &= \frac{\rho_{m}^{N}(r,\theta;\phi_{N}) - \rho_{m}^{x\mathrm{pol}}(r,\theta)\cos[\phi]}{\sin[\phi]} \nonumber\\
&= A(r)\frac{\cos\left[m(\theta-\phi_{N})\right] - \cos\left[m\theta\right]\cos[\phi]}{\sin[\phi]}.\label{Eq:gypol}
\end{align}
By substituting $m\!=\!(Nj \pm 1)$, the first term of the numerator in Eq.~\ref{Eq:gypol}, $\cos[m(\theta-\phi_{N})]$, can be transformed to
\begin{align}
\cos[m(\theta-\phi_{N})] &= \cos[m\theta \mp \phi_{N}],\label{Eq:gtrans}
\end{align}
because $Nj\phi_{N}\!=\!2\pi j$.
From Eqs.~\ref{Eq:gypol} and \ref{Eq:gtrans}, following expression can be derived for the case $m\!=\!Nj\pm1$, as in Eq.~\ref{Eq:case},
\begin{equation}
\rho^{y\mathrm{pol}}_{m\!=Nj\pm1}(r,\theta)\!=\!\pm A(r)\sin[m\theta]\label{Eq:Gcase}.
\end{equation}
Here, $\rho_{m}^{x\mathrm{pol}}$, $\rho_{m}^{N}$, and calculated $\rho_{m}^{y\mathrm{pol}}$ for the case of $m\!=\!N\pm1$ ($j\!=\!1$) in the system with rotational symmetry of $N\!=\!3$ ($m\!=\!2,4$) and $N\!=\!6$ ($m\!=\!5,7$) are plotted in Figure~\ref{Fig:general} (a).
Note that the triangular and hexagonal frames here, and the square frames in Figure~\ref{Fig:model}, are introduced as typical representations of each rotational symmetry and do not indicate a specific geometry.
Owing to Eqs.~\ref{Eq:case} and \ref{Eq:Gcase} being identical except for the $N$, the angular distribution of the $m$-th order mode excited by the LP light with angle $\phi$ for the $C_{N}$ rotational symmetry can also be expressed as
\begin{align}
&\rho_{m\!=\!Nj\pm1}(r, \theta; \phi)\!=\!\rho^{x\mathrm{pol}}_{m}\left(r, \theta\mp\frac{\phi}{m}\right).
\label{Eq:Gadd}
\end{align}
Considering the rotation of the electric field for CP light, Eq.~\ref{Eq:Gadd} indicates that the rotational charge motion in a generalized rotational system can be described as follows: the charge density distribution of azimuthal periodic modes with $m\!=\!Nj\pm1$-order excited in $C_{N}$ symmetric system rotates by $\pm2\pi/m$ in one oscillation period.

\begin{figure*}[t!]
  \begin{center}
  \includegraphics[width=17.2cm]{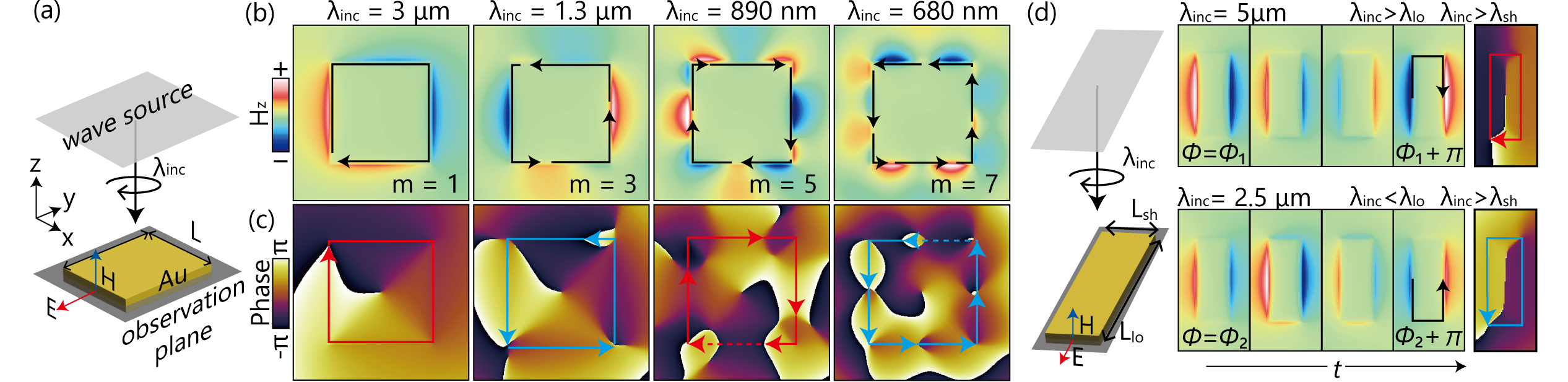}
  \end{center}
  \caption{(a) Schematic of the FDTD simulation model showing the setup with a wave source, an Au plate in the $xy$-plane, and an observation plane. (b) Spatial distributions of the $H_{z}$ component at a representative phase for a square plate with $L\!=\!1~\mathrm{\mu m}$, excited by circularly polarized (CP) light. (c) Spatial phase distributions of the $H_{z}$ component plotted in (b). (d) Spatial distributions over half an oscillation period and spatial phase distributions of $H_{z}$ components for a rectangular nanoantenna at $\lambda\!=\!5~\mathrm{\mu m}$ (top panels) and $2.5~\mathrm{\mu m}$ (bottom panels).}
  \label{Fig:square}
\end{figure*}

Calculated charge density distributions, $\rho_{m}(\theta; \phi\!=\!\pi/4)$ for the $m\!=\!N\pm1$-th modes, are plotted in Figure~\ref{Fig:general} (b).
Similar to the $C_{4}$ system shown in Figure~\ref{Fig:model}, the rotational direction of the charge density distribution for each mode is aligned with that of the electric field for $m\!=\!Nj+1$ and opposite for $m\!=\!Nj-1$.
The amount of the rotation angle is $\phi/m$ in both cases.

It should be noted that the direction and speed of rotation of the charge density obtained here are consistent with one of the accessible OAM values of the far-field scattered field from a rotationally symmetric scatterer excited by CP light~\cite{Lee14NanoP}.
The relationship between the rotational symmetry of the scatterer and the possible OAM values of the far-field scattered field can be understood based on the concept of angular grating, which extends the linear grating to azimuthal periodic polar systems~\cite{Nieminen09JQS, Parker20Optica, Konishi20Optica, Chen14SciRep}.
Our calculations confirm the similarity between the rotational motion of charges within the plasmonic structure and far-field scattered field, both governed by the same restriction of rotational symmetry.


\section{FDTD simulation}
\subsection{Isolated nanoantenna}
To confirm the validity of the model calculations, we evaluated the temporal evolution of the near-field electromagnetic distribution around plasmonic structures using FDTD simulations.
Figure~\ref{Fig:square} (a) shows a schematic of the model used in the FDTD simulation.
A 50 nm thick Au plate with side length $L\!=\!1~\mathrm{\mu}$m is placed in the $xy$-plane, and CP light with wavelength $\lambda_{\mathrm{inc}}$ is normally incident from a total-field scattered-field (TFSF) source located 5 $\mu$m away.
The boundary conditions for the $x$, $y$, and $z$-directions are perfect matched layers (PML), and the minimum mesh width is 10 nm.
The dielectric function of Au is obtained from the Palik database~\cite{Palik}.
The observation plane is set in the middle of the Au layer to record electric field $\bm{E}$ and the magnetic field $\bm{H}$.
All simulations were performed using ANSYS Lumerical (2023 R1).
Since the plasmon waves with a fixed propagation direction have a magnetic field perpendicular to the propagation direction, the excited plasmon mode and their propagation directions can be evaluated from the spatial distribution of the magnetic field.  
Here, we evaluate the $H_{z}$ distribution at the perimeter of the nanoplates to assess the plasmon mode orbiting the outer wall.

The spatial distributions of the $H_{z}$ components at the characteristic wavelengths in the square nanoantenna are plotted in Figure~\ref{Fig:square} (b).
From these distributions, we determined that the resonance orders $m$ are 1, 3, 5, and 7 at wavelengths of 3 $\mu$m, 1.3 $\mu$m, 0.89 $\mu$m, and 0.68 $\mu$m, respectively.
Trajectories of the positive $H_{z}$ components during one oscillation period, evaluated from the time evolution of the magnetic field distribution for each wavelength, are plotted by black arrows in Figure~\ref{Fig:square} (b).
Their direction and length indicate that the rotational directions at adjacent resonance orders are opposite each other, and that the amount of rotation per oscillation period is $2\pi/m$.
These direction and speed of the rotation for each resonance mode are consistent with those obtained from Eq.~\ref{Eq:Gadd}.

Analogous to how the sign and magnitude of the OAMs of a vortex beam can be deduced from the spatial phase distribution of the electromagnetic field, the direction and speed of charge rotation on plasmonic structures can be evaluated from the spatial phase.
Figure~\ref{Fig:square} (c) shows the spatial phase distribution of the $H_{z}$ component for each wavelength plotted in Figure~\ref{Fig:square} (b).
By examining the phase gradient along the perimeter of the square structure, both the direction of the phase gradient and the number of phase cycles per perimeter are observed to be dependent on the resonance order.
The phase gradients from $-\pi$ to $\pi$ along the perimeter are represented by red arrows for clockwise and blue arrows for counterclockwise.
The direction of the phase gradient indicate that the rotational direction of the charges in the $m\!=\!4j+1~(m\!=\!4j-1)$ mode is the same (opposite) as that of the handedness of the excitation CP light.
The region indicated by the dashed line, where the phase gradient is blurred and not distributed from $-\pi$ to $\pi$, is attributable to the contamination from off-resonant adjacent modes with peaks at close frequencies.
In particular, at high frequencies, where the resonance frequencies are close to each other, the electromagnetic field distribution at each wavelength is affected by adjacent modes rotating in opposite directions.

In $C_{N}$ systems where $N\!\geq\!3$, $m$ is uniquely represented by $N$ and $j$, and thus the rotational direction is automatically determined from Eq.~\ref{Eq:Gadd} for each resonance order.
However, for a $C_{2}$ symmetric system, there are two ways to represent $m$, and the direction of charge rotation is not determined solely by rotational symmetry.
For instance, the $m\!=\!1$ case in the $C_{2}$ symmetric system can be represented by $m\!=\!1\!=\!2\cdot0+1$ for $j\!=\!0$ and $m\!=\!1\!=\!2\cdot1-1$ for $j\!=\!1$; consequently, the charge density can rotate either to the clockwise or counter-clockwise.
This is because $\rho_{m}^{y\mathrm{pol}}$ is independent of $\rho_{m}^{x\mathrm{pol}}$ in $C_{2}$ symmetric systems, unlike other rotationally symmetric systems with $N\!\geq\!3$.
Using Eq.~\ref{Eq:case}, the charge density $\rho_{m}^{y\mathrm{pol}}(r,\theta)$ for the $y$-polarized excitation in the above instance is estimated to be $+A(r)\sin\theta$ for $j\!=\!0$ and $-A(r)\sin\theta$ for $j\!=\!1$.

Such conditions can be designed by the geometry of the structure, or by the phase delay of the plasmonic charge motion owing to the difference between the resonance wavelength and the external field.
In the optical resonance of rectangular nanoantennas, the long and short axes, having different resonance wavelengths, $\lambda_{\mathrm{lo}}$ and $\lambda_{\mathrm{sh}}$, exhibit distinct phase delays relative to the external field~\cite{Oshikiri21ACSNano}.
When the external field wavelength ($\lambda_{\mathrm{inc}}$) is much longer than both $\lambda_{\mathrm{lo}}$ and $\lambda_{\mathrm{sh}}$, the resonances on both axes are in phase with the external field.
Conversely, if $\lambda_{\mathrm{inc}}$ lies between $\lambda_{\mathrm{lo}}$ and $\lambda_{\mathrm{sh}}$, the resonances on the short axis oscillate in phase, whereas those on the long axis oscillate out of phase.

The $H_{z}$ components of a rectangular nanoantenna excited by CP light are plotted in Figure~\ref{Fig:square} (d), over a half oscillation period.
The initial phase ($\phi_{1}, \phi_{2}$) is defined where the positive $H_{z}$ is localized on the left side of the rectangle at each wavelength. The trajectories of positive $H_{z}$ components for the half oscillation period are plotted with black arrows in the fourth image.
The lengths of the long and short axes of the rectangle are 1.2 and 0.4 $\mu$m, respectively, and the evaluated resonance wavelengths are $\lambda_{\mathrm{lo}}\!=\!3.6~\mu$m and $\lambda_{\mathrm{sh}}\!=\!1.8~\mu$m.
At $\lambda_{\mathrm{inc}}\!=\!5~\mu$m, which is a longer wavelength than both $\lambda_{\mathrm{lo}}$ and $\lambda_{\mathrm{sh}}$, the $H_{z}$ distribution rotates clockwise along the perimeter of the rectangle, which is the same as the handedness of the CP light (top panels).
In contrast, at $\lambda_{\mathrm{inc}}\!=\!2.5~\mu$m, which is located between $\lambda_{\mathrm{lo}}$ and $\lambda_{\mathrm{sh}}$, the direction is reversed and counterclockwise rotation is observed (bottom panels).
The phase distribution of $H_{z}$ shown in right panels of Figure~\ref{Fig:square} (d) clearly indicates phase gradients in opposite directions.

These results confirm that the rotational behavior of charges in isolated plasmonic nanoantennas is determined by the rotational symmetries of the structure and the resonance mode, as calculated using Eq.~\ref{Eq:Gadd}.
In particular, the results for the rectangular nanoantenna exhibit that the geometric design of the plasmonic structure allows for the selective excitation of rotational charge motion.

\begin{figure*}[t]
  \begin{center}
  \includegraphics[width=17.2cm]{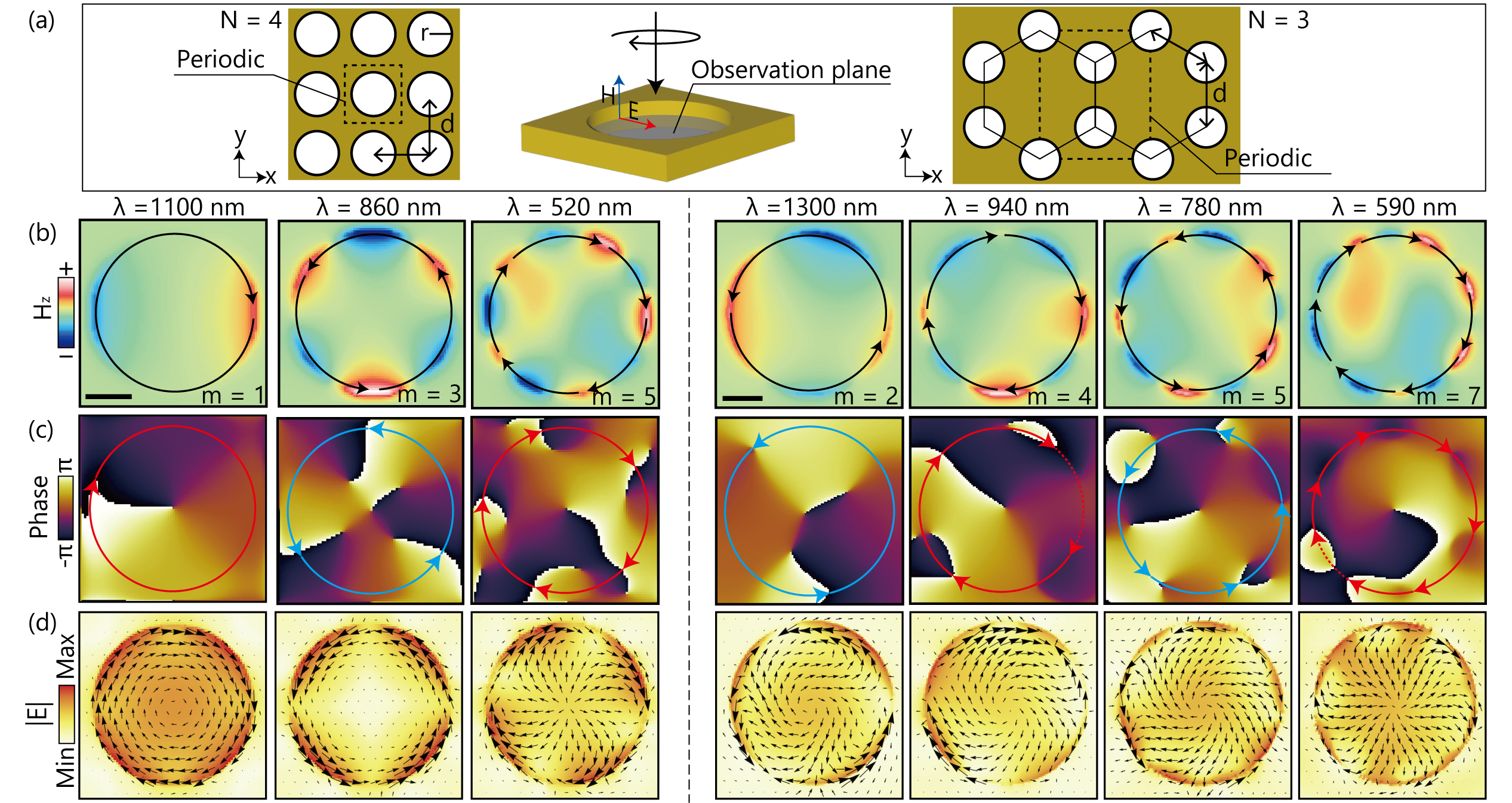}
  \end{center}
  \caption{(a) Schematic of the lattice structures used in the FDTD simulations. (b–d) Spatial distribution of the (b) $H_{z}$ components at representative phase, (c) phase distributions of the $H_{z}$ components, and (d) averaged intensity distributions of the electric field $|E|$ (color map) and Poynting vectors (black arrows). Calculations were performed with $r\!=\!350$ nm and $d\!=\!800$ nm for square lattices, and $r\!=\!400$ nm and $d\!=\!1200$ nm for hexagonal lattices. The scale bars in (b) represent 200 nm.}
  \label{Fig:lattice}
\end{figure*}
\subsection{Periodic lattice}
As the only requirement for Eq.~\ref{Eq:Gadd} is rotational symmetry, the discussion of the rotational behavior of charge densities is applicable to systems other than isolated plates, as long as they are rotationally symmetric.
Circular structures, which generally allow only dipole mode for normally incident plane wave excitation because of their infinite rotational symmetry~\cite{Lan17SciRep,Sakai16SciRep}, can exhibit multipole modes when their symmetry is restricted by an array structure~\cite{Yi17OptCom,Hakala17NC}.
In other words, the multipole modes excited in the periodically arranged structure arise from the external rotational symmetry introduced by the arrangement.
Therefore, the multipole modes excited by CP light in these periodic structures should also exhibit rotational behavior determined by the rotational symmetries.

Here we introduced $C_{4}$ and $C_{3}$ symmetry by arranging circular apertures in square and hexagonal lattice arrays, respectively, as illustrated in Figure~\ref{Fig:lattice}(a).
As with the isolated plate, the resonance order and rotational direction of the excited plasmon mode can be evaluated from the $H_{z}$ component on the inner wall surface.
From the simulation model for the isolated plate used in the previous section, the boundary condition in the $xy$ direction is changed to a periodic boundary, and the TFSF wave source is replaced with a plane wave source.
The lattice structure comprises a 50 nm thick Au thin film in vacuum, featuring an array of holes with radius $r$ and lattice constant $d$.
The minimum mesh size is reduced to 5 nm to minimize the effect of unintended $C_{4}$ symmetry introduced by the Yee lattice in the FDTD simulation.

Figure~\ref{Fig:lattice} (b) shows the spatial distributions of the $H_{z}$ components at representative phases for different resonance orders. The trajectories of the positive $H_{z}$ components per oscillation cycle are also plotted by black arrows, as in Figure~\ref{Fig:square} (b).
The spatial phase distributions of $H_{z}$ components are plotted in Figures~\ref{Fig:lattice} (c), and the phase gradient along the inner wall of the circular aperture is indicated by red and blue arrows.
The rotation angle indicated by each arrow is approximately $2\pi/m$ and is opposite to the incident CP light only when $m\!=\!Nj-1$. This observation is consistent with Eq.~\ref{Eq:Gadd}.

It is noteworthy that the $m\!=\!5$ mode corresponds to $N+1$ for $N\!=\!4$ and $2N-1$ for $N\!=\!3$, indicating that multipole modes of the same order rotate in opposite directions on square and hexagonal lattices.
These calculations demonstrate that the rotational symmetry introduced externally by the periodic array allows us to manipulate the uniform rotational motion of charges in a wide range of plasmonic structures with a high degree of freedom using spatially homogeneous CP light.

One might think that the observed rotational motion of the charges could be due to shifts in the constructive interference position or the wagon wheel effect resulting from the discrete computation time.
Here, we investigate the spatial distribution of the Poynting vector $\bm{P}\!=\!\mathrm{Re}[\bm{E^*}\times\bm{H}]/2$, which represents the energy flow.
Figure~\ref{Fig:lattice} (d) shows the in-plane components of the Poynting vectors, as black arrows on the color map of the averaged electric field intensity. 
The orientation of the Poynting vectors along the circumference corroborates that the clockwise and counterclockwise charge motions actually possess an energy flow consistent with their rotation direction.
These energy flow indicate that the rotation of charges excited by CP light can be regarded as a surface current orbiting the plasmonic structure. 
The non-uniform intensity distribution and partially retrograde pointing vectors (e.g., the right side of $\lambda\!=\!940$ nm for $N\!=\!3$) are attributed to the contamination from off-resonant adjacent modes.

Manipulating the rotational motion of charge distributions in periodic plasmonic nanostructures shows that uniform control of rotational charge motion in large-area structures suitable for experimental observation is achievable. This capability offers a variety of potential applications, such as the observation of the inverse Faraday effect~\cite{Alma24NanoP,Parchenko24APR, Karakhanyan22PRB, Cheng20NatP, Mou23AMT}, designing the resonant interaction between magnetic resonance antennas~\cite{Ichiji22OL}, manipulation of the photocurrent~\cite{Li15NC, Wang24AMI, Mou24NL} and spin currents~\cite{oue2020effects,oue2020electron,oue2020optically,ukhtary2020surface,ukhtary2021spin,tian2022switching,bekshaev2022spin}, and control of the SAM associated with the plasmonic near-field~\cite{Triolo17ACSP, Shi21PNAS,Ichiji23PRA}.

The spatial phase distributions similar to those of vortex beams with higher-order OAMs observed in Figure~\ref{Fig:square} (c) and~\ref{Fig:lattice} (c) suggest the possibility of experimental investigations of light-matter interactions involving OAM~\cite{Porfirev23PQE,Ni23NatP, Rouxel22NP,Jain24NC,Cao23ACSP}.
Furthermore, our calculations indicate that the rotational charge motions on plasmonic structures excited by CP light exhibit opposite directions between adjacent modes among the multipole resonance modes that can be excited in each structure.
This is important for explaining certain complex spatial distributions and non-intuitive temporal variations of electromagnetic fields often observed in electromagnetic simulations using CP excitation light~\cite{Ichiji24PRB}, and aids in the calculation of the optical properties of enhanced electromagnetic fields, such as SAM~\cite{Triolo17ACSP} and optical chirality~\cite{Schaeferling12PRX}.

Although we limited the structures in each calculation to one type for simplicity, more complex rotational behavior of the charges can be designed by combining different structures. For instance, charges can be rotated in opposite directions on the same plane by using structures of different sizes that share the same rotational symmetry (see appendix).
Designing geometries other than regular polygons, such as swastikas, would facilitate a more efficient excitation of specific rotational behaviors. 
The actual charge density distribution in nanopolygon plates often does not exhibit the perfect azimuthal periodicity as simplified in this study ~\cite{Schmidt14NL,Pellarin16ACSN}. Defining basis vectors based on modal analysis that considers the geometry of each structure to evaluate such complex distributions is our future work~\cite{Boudarham12PRB}.

\section{Conclusion}
In conclusion, we numerically investigated the rotational charge behavior of nanoscale plasmonic structures excited by CP light. By simplifying the spatial distribution of the multipole plasmon modes, we formulated the manner in which the rotational symmetry of the plasmonic structure and that of the plasmon mode affect the charge motion.
We showed that the direction and speed of charge rotation on the plasmonic structure could be controlled using spatially homogeneous CP light, by appropriately selecting these two rotational symmetries.

The FDTD simulations validated the charge behavior predicted by the analytical model for both single nanoplates and periodically arranged hole structures, thus demonstrating that the rotational charge motion could be effectively managed by plane CP light across a wide range of plasmonic structures.
The findings on the rotational behavior of charges summarized in this study are expected to greatly aid in the experimental verification of light-matter interactions involving angular momentum.

\section*{Acknowledgments}
This work was supported by a Grant-in-Aid for JSPS Fellows (No.JP23KJ0355), a Grant-in-Aid for Scientific Research (A) (No. JP20H00325), and a Grant-in-Aid for Challenging Exploratory Research (No.JP20K20560).
D.O. is supported by JSPS Overseas Research Fellowship, by the Institution of Engineering and Technology (IET), and by Funda\c{c}\~ao para a Ci\^encia e a Tecnologia and Instituto de Telecomunica\c{c}\~oes under project UIDB/50008/2020.

\vspace{0.3cm}
\appendix
\renewcommand\thefigure{S\arabic{figure}} 
\setcounter{figure}{0}  

\section{Simulation results of combined structure}
\begin{figure}[h!]
\centering
\includegraphics[width=8.6 cm]{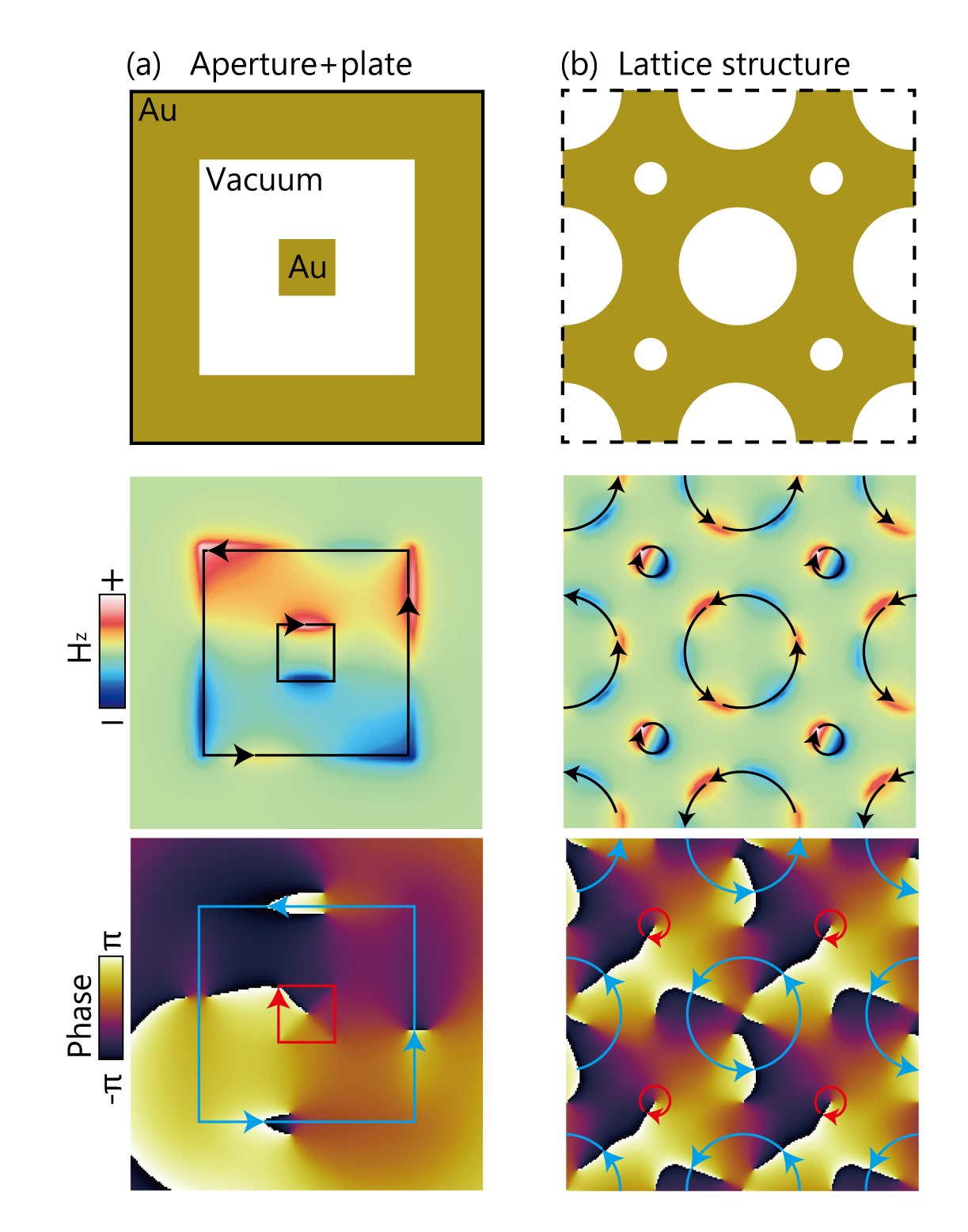}
\caption{Simulation results for plasmonic structures combining elements of different sizes sharing the same rotational symmetry, excited with circularly polarized light. (a) a combination of a large square aperture ($L = 600$ nm) with a small square plate ($L = 150$ nm) in the center. (b) a combination of $C_{4}$ symmetric lattice arrays of circular aperture structures with radii of 200 nm and 50 nm.}
\label{Fig:sp1}
\end{figure}

\newpage
\bibliography{apssamp}

\end{document}